\documentstyle[12pt]{article}
\newcommand{\be}{\begin{equation}}
\newcommand{\ee}{\end{equation}}
\newcommand{\bea}{\begin{eqnarray}}
\newcommand{\eea}{\end{eqnarray}}

\newcommand{\eps}{\varepsilon}

\newcommand{\half}{{\scriptstyle{{1\over 2}}}}

\newcommand{\real}{\relax{\rm I\kern-.18em R}}
\newcommand{\Tr}{{\rm Tr}}
\newcommand{\tr}{{\rm tr}}

\newcommand{\cA}{{\cal A}}

\newcommand{\cP}{{\cal P}}
\newcommand{\cO}{{\cal O}}

\def\pl{{{\cal P}_\infty}}

\def\myre{{\rm Re}}
\def\ba{\begin{array}}
\def\ea{\end{array}}

\begin{document}
\vskip-1cm
\hfill FTUAM-99-12; IFT-UAM/CSIC-99-16
\vskip0mm
\hfill INLO-PUB-8/99
\vskip5mm
\begin{center}
{\LARGE{\bf{\underline{Weyl-Dirac zero-mode for calorons}}}}\\
\vspace*{10mm}{\large
Margarita Garc\'{\i}a P\'erez$^{(a)}$, Antonio Gonz\'alez-Arroyo$^{(a,b)}$,\\
Carlos Pena$^{(a)}$ and Pierre van Baal$^{(c)}$\\}
\vspace*{8mm}
($a$) Departamento de F\'{\i}sica Te\'{o}rica C-XI,\\
Universidad Aut\'{o}noma de Madrid,\\
28049 Madrid, Spain.\\
\vspace*{3mm}
($b$) Instituto de F\'{\i}sica Te\'{o}rica C-XVI,\\
Universidad Aut\'{o}noma de Madrid,\\
28049 Madrid, Spain.\\
\vspace*{3mm}
($c$) Instituut-Lorentz for Theoretical Physics,\\
University of Leiden, PO Box 9506,\\
NL-2300 RA Leiden, The Netherlands.
\end{center}
\vspace*{5mm}{\narrower{\noindent
\underline{Abstract:} We give the analytic result for the fermion
zero-mode of the $SU(2)$ calorons {\em with non-trivial holonomy}. It is 
shown that the zero-mode is supported on {\em only one} of the constituent
monopoles. We discuss some of its implications.}\par}
\section{Introduction}
In this paper we give the exact expression for the $SU(2)$ fermion zero-mode 
in the field of the infinite volume caloron with non-trivial holonomy and unit 
charge. Study of the gauge field configurations had, somewhat surprisingly, 
revealed that at non-trivial holonomy calorons have two BPS monopoles ($N$ for
$SU(N)$) as their constituents~\cite{KrvB,KrvN,LeeL}. For the 
Harrington-Shepard~\cite{HaSh} solution with trivial holonomy this is hidden, 
because one of the constituents is massless (it can be removed by a singular 
gauge transformation to show that the caloron for a large scale parameter 
becomes a single BPS monopole~\cite{Rossi}). We find for calorons with 
well-separated constituents, that the fermion zero-mode is entirely supported 
on one of them. In itself it is not surprising that the zero-mode is correlated
to the monopole constituents. Independently this observation was recently also 
made for gluino zero-modes in the context of supersymmetric gauge 
theories~\cite{SUSY}. Gluinos are in the adjoint representation of the gauge 
group, such that there are four zero-modes, that can be split in pairs 
associated to each of the two constituents~\cite{SUSY}. However, for the Dirac 
fermion there is only one zero-mode. To understand the ``affinity'' of the 
zero-mode to only one of the two monopoles, we will analyse in some detail 
what distinguishes them.

Calorons are characterised by the (fixed) holonomy~\cite{KrvB,GPY}. In the
gauge in which $A_\mu(x)$ is periodic, this holonomy is given by
\be
\pl=\lim_{|\vec x|\rightarrow\infty} {\rm P}\exp\left(\int_0^\beta A_0(t,\vec x)
dt\right)\equiv\exp(2\pi i\vec\omega\cdot\vec\tau).
\ee
Solutions are simplest in the so-called ``algebraic'' gauge, for which
\be
A_\mu(t+\beta,\vec x)=\pl A_\mu(t,\vec x)\pl^{-1},\quad
\Psi^\pm_z(t+\beta,\vec x)=\pm\pl\Psi_z^\pm(t,\vec x).
\ee

We will generalise the problem of finding the fermion zero-mode in the field 
of the caloron with non-trivial holonomy, by adding a curvature free Abelian 
field, which forms the basis for the Nahm transformation~\cite{Nahm}. Why
this is useful will be evident from the construction. The equation to be 
solved is
\be
\bar D_z\Psi_z(x)\equiv\bar\sigma_\mu(\partial_\mu+A_\mu(x)-2\pi iz_\mu)\Psi_z
(x)=0,\label{eq:Weyl}
\ee
with $\bar\sigma_\mu=\sigma_\mu^\dagger=(1,-i\vec\tau)$ and $\tau_a$ the 
Pauli matrices. For calorons, defined on $R^3\times S^1$, i.e. at finite 
temperature $1/\beta$, one can choose $z_1=z_2=z_3=0$ (the plane-wave factor 
$\exp(2\pi i\vec z\cdot\vec x)$ does not affect the boundary conditions or
the normalisation of the zero-mode and can be used to remove the $\vec z$
dependence). But $z=z_0$ will be arbitrary (it has a dual period of $1/\beta$). 
The zero-modes are represented as two-component spinors in the (chiral) Weyl 
decomposition for massless Dirac fermions. 

\section{ADHM construction}
The construction of the zero-mode is best done in the ADHM 
formalism~\cite{ADHM}. We will be brief in reviewing this formalism, further
details can be found in ref.~\cite{KrvB,Temp,Osb,CoGo}. In general the ADHM 
construction involves an operator $\Delta(x)$ (the ``dual'' of 
eq.~(\ref{eq:Weyl})), whose normalised zero-mode, $\Delta^\dagger(x)v(x)=0$, 
gives the gauge field as $A_\mu(x)=v^\dagger(x)\partial_\mu v(x)$. 
For $SU(2)$  instantons of charge $k$ one has $\Delta^\dagger=(\lambda^\dagger,
B^\dagger-x^\dagger)$, with $x^\dagger=x_\mu\bar\sigma_\mu$, $\lambda$ a $k$ 
dimensional row vector and $B$ a $k\times k$ symmetric matrix, all with 
values in the quaternions ($\lambda^\ell_{iI}$, $B^{\ell,m}_{IJ}=
B^{m,\ell}_{IJ}$, with $\ell,m=1,\cdots,k$ ``charge'' and $I,J=1,2$ spinor 
indices, whereas $i=1,2$ is a colour index). Introducing the row vector 
$u^\dagger(x)\equiv\lambda(B-x)^{-1}$ and the scalar (real quaternion) 
$\phi(x)=1+u^\dagger(x)u(x)$, it can be shown~\cite{Temp,Osb,CoGo} that the 
instanton gauge field and the $k$ zero-modes are given by
\be
A_\mu(x)=\phi^{-1}(x){\rm Im}\left(u^\dagger(x)
\partial_\mu u(x)\right),\quad \Psi^\ell_{iJ}(x)=
\pi^{-1}\phi^{-\half}(x)(u^\dagger(x)f_x)^\ell_{iI}\,\eps_{IJ},
\label{eq:zm}
\ee
where $f_x$ is the matrix inverse, or Green's function,
\be
f_x=\left((B-x)^\dagger(B-x)+\lambda^\dagger\lambda\right)^{-1}.
\ee
Essential in the ADHM construction is that $\lambda$ and $B$ satisfy a 
quadratic constraint, which is equivalent to $f_x$ being a symmetric matrix 
whose imaginary quaternion components vanish, $(f_x)^{\ell,m}_{IJ}\equiv
f_x^{\ell,m}\delta_{IJ}$. From this alone it can be proven that the gauge 
field is self-dual and that the $\Psi^\ell(x)$ are zero-modes, $\bar D\Psi^\ell
(x)\equiv\bar\sigma_\mu(\partial_\mu+A_\mu(x))\Psi^\ell(x)=0$. Its proper 
normalisation and the topological charge are read off from the remarkable 
results~\cite{Temp,Osb,CoGo}
\be
\Psi^\ell(x)^\dagger\Psi^m(x)=-(2\pi)^{-2}\partial_\mu^2 f_x^{\ell,m},
\quad 
\Tr F_{\mu\nu}^{\,2}(x)=-\partial_\mu^2\partial_\nu^2\log\det f_x ,
\label{eq:mir}
\ee
using $\lim_{|x|\rightarrow\infty}f_x^{\ell,m}=\delta^{\ell,m}|x|^{-2}$.
Before addressing the explicit form of these expressions for the caloron
with non-trivial holonomy, we perform one further simplification (for the
details follow eqs.~(21-29) in ref.~\cite{KrvB}, see also ref.~\cite{Temp}), 
\be
A_\mu(x)=\half\phi(x)\partial_\nu\left(\lambda\bar\eta_{\mu\nu}f_x
\lambda^\dagger\right),\quad \Psi^\ell_{iJ}(x)=(2\pi)^{-1}\phi^\half(x)
\partial_\mu\left(\lambda f_x\bar\sigma_\mu^{}\right)^\ell_{iI}\eps_{IJ},
\ee
where the anti-selfdual 't~Hooft tensor $\bar\eta$ is defined by 
$\bar\eta^i_{0j}=-\bar\eta^i_{j0}=\delta_{ij}$ and $\bar\eta^i_{jk}=
\eps_{ijk}$ (furthermore $\eps_{12}=1$ and with our conventions of 
$t=x_0$, $\eps_{0123}=-1$).

The caloron with non-trivial holonomy is found by imposing boundary conditions
to compactify time to a circle $u^{\ell+1}(t+\beta,\vec x)=u^\ell(t,\vec x)
\cP_\infty^\dagger$, which is easily seen to give the correct boundary 
conditions for the gauge field. For the general form of $\lambda$ and $B$ 
which respect this symmetry, see ref.~\cite{KrvB}. Note that now the index 
$\ell$ runs over the set of all integers; the $R^4$ configuration with these 
boundary conditions has infinite topological charge (unit topological charge
per time-period). To obtain the zero-mode with the appropriate boundary 
condition we note that with eq.~(\ref{eq:zm})
\be
\hat\Psi_z(x)\equiv\sum_\ell e^{-2\pi i\ell\beta z}\Psi^\ell(x),
\ee
satisfies the boundary condition $\hat\Psi_z(t+\beta,\vec x)=e^{-2\pi i\beta z}
\pl\hat\Psi_z(t,\vec x)$ and satisfies $\bar D\hat\Psi_z(x)=0$ for all $z$.
The general solution of the Weyl equation, eq.~(\ref{eq:Weyl}), with both 
periodic and anti-periodic boundary conditions is now easily found (for
simplicity we put $\beta=1$) 
\be
\Psi^+_z(x)=e^{2\pi izt}\hat\Psi_z(x),\quad
\Psi^-_z(x)=e^{2\pi izt}\hat\Psi_{z+1/2}(x).\quad
\ee
In particular $\Psi^-(x)=\Psi^-_0(x)=\hat\Psi_{1/2}(x)$ is the, for 
finite temperature, physically relevant chiral zero-mode in the background 
of a caloron, whereas $\Psi^+(x)=\Psi^+_0(x)=\hat\Psi_0(x)$ is relevant for
compactifications.
\section{Nahm-Fourier transformation}
The interpretation of the ``charge'' index as a Fourier index, as suggested
by the construction of the caloron zero-mode, has been essential for solving
the quadratic ADHM constraint in the presence of non-trivial holonomy.
It maps the ADHM construction to the Nahm formalism, in which furthermore $f_x$ 
is solved in terms of a quantum mechanics problem on the circle ($z\in[0,1]$) 
with a piecewise constant potential and delta function singularities 
determined by the holonomy~\cite{KrvB}. The relevant quantities involved are
\be
\hat f_x(z,z')=\sum_{\ell,m}f_x^{\ell,m}e^{2\pi i(\ell z-mz')},\quad
\hat\lambda(z)=\sum_\ell\lambda^\ell e^{-2\pi i\ell z},
\ee
where matrix multiplication is replaced by convolution in the usual sense.
The solution of the ADHM constraint implies that $\hat\lambda(z)$ is the 
sum of two delta functions. Together with the explicit expression for the 
Green's function $\hat f_x(z,z')$ as given in eqs.~(47-49) of ref.~\cite{KrvB}, 
the zero-mode reads
\be
\hat\Psi_z(x)=(2\pi)^{-1}\phi^\half(x)\partial_\mu\left(\int_0^1dz'
\hat\lambda(z')\hat f_x(z',z)\bar\sigma_\mu\right)_{iI}\eps_{IJ},
\ee
cmp. eq.~(\ref{eq:zm}). Whereas eq.~(\ref{eq:mir}) yields
\be
\hat\Psi_{z'}^\dagger(x)\hat\Psi_z(x)=-(2\pi)^{-2}\partial_\mu^2\hat f_x(z',z).
\ee
\section{Explicit expressions}
Using the classical scale invariance to put $\beta=1$, one has~\cite{KrvB}
\be
s(x)=-\half\Tr F_{\mu\nu}^{\,2}(x)=-\half\partial_\mu^2\partial_\nu^2\log
\psi(x),\quad\psi(x)=\hat\psi(\vec x)-\cos(2\pi t),
\quad\hat\psi(\vec x)=\half\tr(\cA_2 \cA_1),
\ee
where
\be
\cA_m\equiv\frac{1}{r_m}\left(\!\!\!\ba{cc}r_m\!\!&|\vec y_m\!\!-
\!\vec y_{m+1}|\\0\!\!&r_{m+1}\ea\!\!\!\right)\left(\!\!\!\ba{cc}c_m\!\!&
s_m\\s_m\!\!&c_m\ea\!\!\!\right).
\ee
Noting that $r_{3}\equiv r_1$ and $\vec y_{3}\equiv\vec y_1$ we defined
$r_m=|\vec x-\vec y_m|$, with $\vec y_m$ the position of the $m^{\rm th}$ 
constituent monopole, which can be assigned a mass $8\pi^2\nu_m$, where 
$\nu_1=2\omega$ and $\nu_2=2\bar\omega\equiv1-2\omega$. Furthermore, 
$c_m\equiv\cosh(2\pi\nu_m r_m)$, $s_m\equiv\sinh(2\pi\nu_m r_m)$ and
$\pi\rho^2=|\vec y_2-\vec y_1|$. 

New is the result for the zero-mode density 
\be
|\Psi^-(x)|^2=-(2\pi)^{-2}\partial_\mu^2 \hat f_x(\half,\half),\quad
|\Psi^+(x)|^2=-(2\pi)^{-2}\partial_\mu^2 \hat f_x(0,0),
\ee
defined by~\cite{KrvB}
\bea
\hat f_x(\half,\half)=\frac{\pi}{r_1r_2\psi(x)}\left\{s_2[r_1c_1+\pi\rho^2s_1]
+r_2s_1+\frac{c_2\!-\!1}{r_2}[\pi\rho^2r_1c_1+\half(r_1^2+r_2^2+\pi^2\rho^4)s_1]
\right\},\nonumber\\
\hat f_x(0,0)=\frac{\pi}{r_1r_2\psi(x)}\left\{s_1[r_2c_2+\pi\rho^2s_2]
+r_1s_2+\frac{c_1\!-\!1}{r_1}[\pi\rho^2r_2c_2+\half(r_1^2+r_2^2+\pi^2\rho^4)s_2]
\right\}.\label{eq:norm}
\eea

By a suitable combination of a constant gauge transformation, spatial rotation 
and translation we can arrange both $\vec\omega=(0,0,\omega)$ and the 
constituents at $\vec y_1=(0,0,\nu_2\pi\rho^2)$ and 
$\vec y_2=(0,0,-\nu_1\pi\rho^2)$. For this choice we find
\bea
A_\mu(x)=\frac{i}{2}\bar\eta^3_{\mu\nu}\tau_3\partial_\nu\log\phi(x)+
\frac{i}{2}\phi(x)\myre\left((\bar\eta^1_{\mu\nu}-i\bar\eta^2_{\mu\nu})
(\tau_1+i\tau_2)\partial_\nu\chi(x)\right),\\
\Psi^-_{1I}(x)=(2\pi)^{-1}\rho\phi^\half(x)\pmatrix{\partial_2+i\partial_1\cr
\partial_0-i\partial_3\cr}\hat f_x(\omega,\half),\quad \Psi^-_{2I}(x)=
\Psi^-_{1J}(x)^*\eps_{JI},\nonumber\\
\Psi^+_{1I}(x)=(2\pi)^{-1}\rho\phi^\half(x)\pmatrix{\partial_2+i\partial_1\cr
\partial_0-i\partial_3\cr}\hat f_x(\omega,0),\quad \Psi^+_{2I}(x)=
\Psi^+_{1J}(x)^*\eps_{JI},
\eea
with $\phi^{-1}(x)=1-\frac{\pi\rho^2}{\psi(x)}\left(\frac{s_1c_2}{r_1}
+\frac{s_2c_1}{r_2}+\frac{\pi\rho^2 s_1s_2}{r_1r_2}\right)$ and $\chi(x)=
\frac{\pi\rho^2}{\psi(x)}\left(e^{-2\pi it}\frac{s_1}{r_1}+\frac{s_2}{r_2}
\right)e^{2\pi i\nu_1 t}$, and
\bea
\hat f_x(\omega,\half)=\frac{\pi e^{\pi i\nu_1 t}}{r_1r_2\psi(x)}
\left\{\left(e^{\pi it}r_1\!+\!e^{-\pi it}[\pi\rho^2s_1\!+\!r_1c_1]\right)
\sinh(\pi r_2\nu_2)\!+\!e^{-\pi it}r_2s_1\cosh(\pi r_2\nu_2)\right\},\nonumber\\
\hat f_x(\omega,0)=\frac{\pi e^{-\pi i\nu_2 t}}{r_1r_2\psi(x)}
\left\{\left(e^{-\pi it}r_2+e^{\pi it}[\pi\rho^2s_2\!+\!r_2c_2]\right)
\sinh(\pi r_1\nu_1)+e^{\pi it}r_1s_2\cosh(\pi r_1\nu_1)\right\}.
\eea
\section{Properties of the zero-mode}
\begin{figure}[htb]
\vspace{4.5cm}
\includegraphics{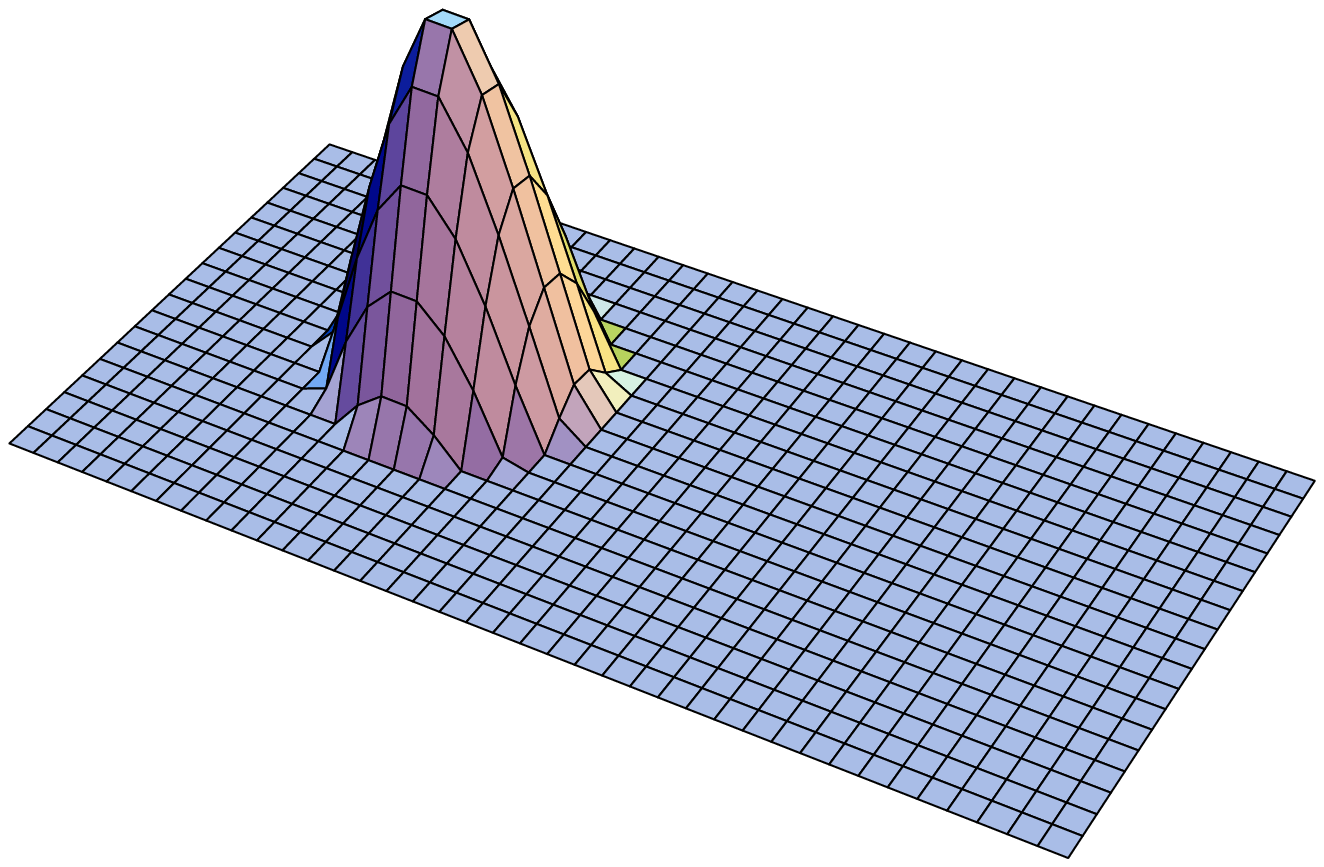}
\includegraphics{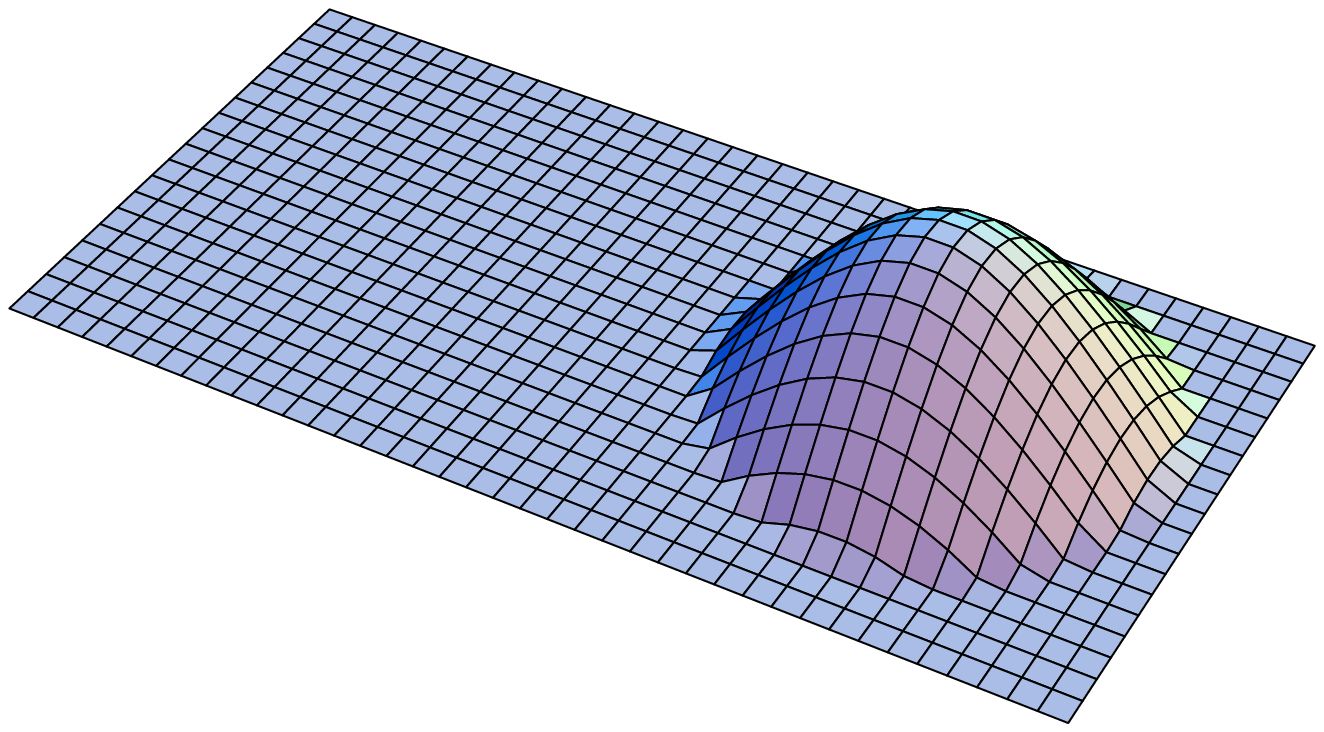}
\includegraphics{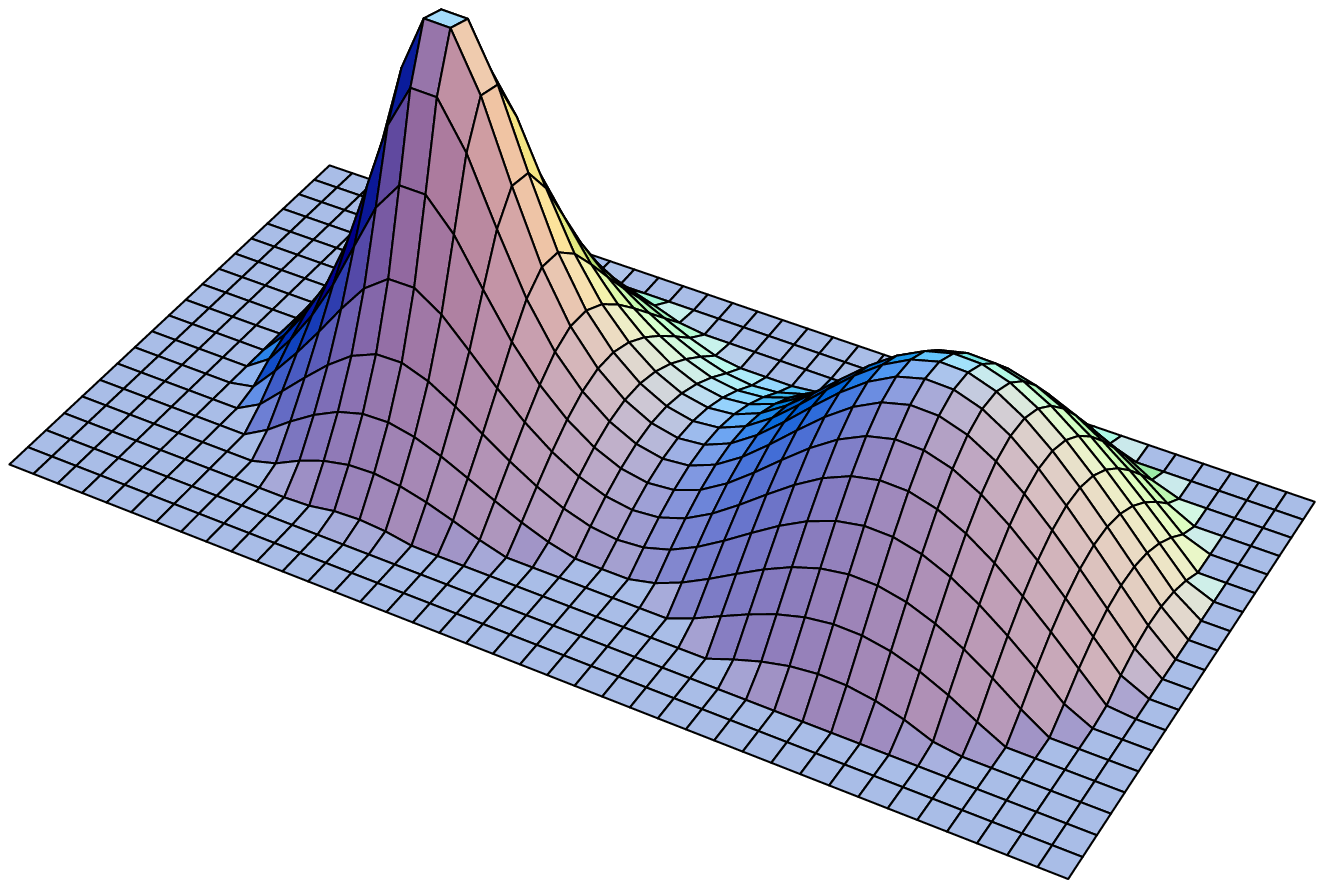}
\caption{For the two figures on the sides we plot on the same scale the 
logarithm of the zero-mode densities (cutoff below $1/e^5$) for $\omega=1/8$ 
(left $\Psi^-$ / right $\Psi^+$) and $\omega=3/8$ (right $\Psi^-$ / left 
$\Psi^+$), with $\beta=1$ and $\rho=1.2$. In the middle figure we show for
the same parameters (both choices of $\omega$ give the same action density)
the logarithm of the action density (cutoff below $1/2e^2$).}
\label{fig:zm}
\end{figure}

The gauge field has a symmetry under the anti-periodic gauge transformation 
$g(x)=\exp(\pi it\hat\omega\cdot\vec\tau)$, which changes the sign of the 
holonomy, $\pl\rightarrow -\pl$, or $\omega\leftrightarrow\bar\omega=\half-
\omega$. An anti-periodic gauge transformation does, however, not leave 
fermions invariant, and indeed it interchanges $\Psi^+_z(x)$ and $\Psi^-_z(x)$. 
To preserve the special choice of parametrisation presented above, the change 
of sign in the holonomy, which interchanges $\nu_1$ and $\nu_2$, is also 
accompanied by an interchange of the constituent locations. This indeed leaves 
the action density invariant. That the zero-mode clearly distinguishes between 
the two cases becomes evident in the static limit, $\rho\rightarrow\infty$ (or 
equivalently $\beta\rightarrow0$), in which case the zero-mode is completely 
localised on one of the constituent monopoles, as follows from (cmp. 
eq.~(\ref{eq:norm}))
\be
\lim_{\rho\rightarrow\infty}|\Psi^-(x)|^2=-\partial_\mu^2\left\{
\frac{\tanh(\pi r_2\nu_2)}{4\pi r_2}\right\},\quad 
\lim_{\rho\rightarrow\infty}|\Psi^+(x)|^2=-\partial_\mu^2\left\{
\frac{\tanh(\pi r_1\nu_1)}{4\pi r_1}\right\}.
\ee
Under the anti-periodic gauge transformation the anti-periodic zero-mode 
becomes periodic. The {\em new} anti-periodic zero-mode is now completely 
localised on the {\em other} constituent monopole (this is consistent with 
the fact that $\hat f_x(0,0)$ can be obtained from $\hat f_x(\half,\half)$ 
by interchanging $r_1$ and $\nu_1$ with $r_2$ and $\nu_2$). Figure 
\ref{fig:zm} illustrates these issues~\cite{WWW}.

In the gauge where $A_\mu(x)$ is periodic, for large $\rho$ one of the 
constituent monopole fields is completely time independent, whereas the other 
one has a time dependence that would result from a full rotation along the 
axis connecting the two constituents~\cite{KrvB}. This is read off from
\bea
&&\hskip-6mm A_\mu^{\rm per}=\frac{i}{2}\bar\eta^3_{\mu\nu}\tau_3\partial_\nu
\log\phi+\frac{i}{2}\phi\myre\left((\bar\eta^1_{\mu\nu}-i\bar\eta^2_{\mu\nu})
(\tau_1+i\tau_2)(\partial_\nu+4\pi i\omega\delta_{\nu,0})\tilde\chi\right)\!
+\!\delta_{\mu,0}2\pi i\omega\tau_3,\\
&&\hskip-6mm\tilde\chi=e^{-4\pi it\omega}\chi=\frac{4\pi\rho^2}{(r_2+r_1+\pi
\rho^2)^2}\left\{r_2e^{-2\pi r_2\nu_2}e^{-2\pi it}\!+r_1 e^{-2\pi r_1\nu_1}
\right\}\left(1\!+\!\cO(e^{-4\pi\,{\rm min}(r_1\nu_1,r_2\nu_2)})\right).
\nonumber
\eea
(Note that for large $\rho$, $\phi(x)$ becomes time independent.)
This full rotation - which we will call the {\em Taubes-winding} - is 
responsible for the topological charge of the otherwise time independent 
monopole pair~\cite{Taubes}. Under the anti-periodic gauge transformation that 
changes the sign of the holonomy the Taubes-winding is supported by the other 
constituent. It has not gone unnoticed that the anti-periodic fermion 
zero-mode is precisely localised on the monopole constituent that carries 
the Taubes-winding. Another way to distinguish the two constituent monopoles 
is by inspecting the Polyakov loop values at their centers. One finds -1 for 
the monopole line with the Taubes-winding and +1 for the other monopole line 
(this is correlated to the vanishing of the would-be Higgs field), see the 
appendix of ref.~\cite{GGMV}. For trivial holonomy, $\omega=0$, the Polyakov
loop is indeed -1 at the center of the Harrington-Shepard~\cite{HaSh} caloron.
Its zero-mode, constructed before in ref.~\cite{GPY,GoSi}, agrees with the 
results found here.

The association of the zero-mode with the constituent that carries the 
Taubes-winding lends considerable support for the role of the monopole loops 
with Taubes-winding in QCD for chiral dynamics~\cite{KrvB}. The precise 
embedding of these straight finite temperature monopole loops as curved 
monopole loops in flat space remains a non-trivial and challenging problem. 
Although it may seem contradictory to expect the zero-mode with anti-periodic 
boundary conditions to be the relevant one, one should not forget that for a 
curved monopole loop the spin frame makes also one full rotation due to the 
{\em bending} of the loop, thereby most likely providing the compensating sign 
flip.
\section*{Acknowledgements}
We are grateful to Maxim Chernodub, Arjan Keurentjes, Valya Khoze and Tamas 
Kov\'acs for useful discussions and correspondence. A. Gonzalez-Arroyo and 
C. Pena acknowledge financial support by CICYT  under grant AEN97-1678.
M. Garc\'{\i}a P\'erez acknowledges financial support by CICYT.

\end{document}